\documentclass[prl,aps,preprint,showpacs]{revtex4-1}
\usepackage{graphicx}
\usepackage{amsmath}
\usepackage[table]{xcolor}

\begin{document}

\title{Large-gap quantum spin Hall insulators in tin films}

\author{Yong \surname{Xu}$^{1,2}$}
\author{Binghai \surname{Yan}$^{3}$}
\author{Hai-Jun \surname{Zhang}$^{1}$}
\author{Jing \surname{Wang}$^{1}$}
\author{Gang \surname{Xu}$^{1}$}
\author{Peizhe \surname{Tang}$^{2}$}
\author{Wenhui \surname{Duan}$^{2}$}
\author{Shou-Cheng \surname{Zhang}$^{1,2}$}
\email{sczhang@stanford.edu}

\affiliation{$^1$Department of Physics, McCullough Building, Stanford University, Stanford, California 94305-4045, USA \\
$^2$Institute for Advanced Study, Tsinghua University, Beijing 100084, People's Republic of China \\
$^3$Max Planck Institute for Chemical Physics of Solids, 01187 Dresden, Germany}

\begin{abstract}
The search of large-gap quantum spin Hall (QSH) insulators and effective approaches to tune QSH states is important for both fundamental and practical interests. Based on first-principles calculations we find two-dimensional tin films are QSH insulators with sizable bulk gaps of 0.3 eV, sufficiently large for practical applications at room temperature. These QSH states can be effectively tuned by chemical functionalization and by external strain. The mechanism for the QSH effect in this system is band inversion at the $\Gamma$ point, similar to the case of HgTe quantum well. With surface doping of magnetic elements, the quantum anomalous Hall effect could also be realized.
\end{abstract}

\pacs{73.43.-f, 71.70.Ej, 73.22.-f}


\maketitle

Topological insulators are new states of quantum matter interesting for both fundamental condense-matter physics and material science~\cite{qi2010quantum,hasan2010colloquium,qi2011topological,yan2012topological}. Such states are characterized by an insulating bulk, and metallic surface/edge states protected by time-reversal symmetry. Whereas the surface states of three-dimensional topological insulators are not protected against scattering at any angle other than 180 degrees, the robustness of conducting edge states from backscattering has already been demonstrated in the two-dimensional topological insulator HgTe quantum well~\cite{bernevig2006quantum,konig2007quantum}. This is promising for the realization of conducting channels without dissipation.

The stoichiometric crystals Bi$_2$Se$_3$, Bi$_2$Te$_3$ and Sb$_2$Te$_3$~\cite{Bi2X3-natphy, Bi2Te3-exp-Chen, Bi2Se3-exp-Hasan} offer ideal model systems for the experimental investigation of three-dimensional (3D) TIs. For two-dimensional (2D) TIs or quantum spin Hall (QSH) insulators, HgTe quantum well is a well-established system, but its bulk gap is too small, with the QSH effect observed only at low temperatures (below 10 K)~\cite{konig2007quantum}. Extensive effort has been devoted to search new QSH insulators~\cite{kane2005quantum,murakami2006quantum,liu2011stable,liu2008quantum,knez2011evidence,xiao2011interface,liu2011quantum}, however, desirable materials preferably with large bulk gaps are still lacking.

On the other hand, following the success of graphene, various chemical classes of 2D materials including novel materials initially considered to exist only in the realm of theory
have been synthesized~\cite{novoselov2005two,Butler2013Progress}. Particularly, the 2D group IV honeycomb lattices, that are interesting for electronics~\cite{cahangirov2009two}, have been successively fabricated. For instance, a hydrogenated graphene (graphane)~\cite{elias2009control}, a silicon counterpart of graphene (silicene)~\cite{vogt2012silicene} and a germanium graphane analogue~\cite{bianco2013stability} were experimentally found. Also, ultrathin tin films that are presumably to be buckled monolayer and in a honeycomb lattice were observed by early molecular beam epitaxy experiments~\cite{osaka1994surface,zimmermann1997growth}.

This Letter reports a series of large-gap QSH insulators in functionalized tin films, based on first-principles calculations. The chemical symbol of tin is Sn, originating from the Latin word  "Stanum" for tin, therefore a monolayer of tin film can be called ``stanene'', in analogy with graphene and silicene. Similarly, the hydrogen terminated tin film could be called ``stanane''. These new QSH insulators have extraordinarily large bulk gaps ($\sim$0.3 eV), their QSH states can be effectively tuned by chemical functionalization and by external strain, and their use is benefited from the abundant degree of freedom in the chemical functional group. All these make tin films intriguing for applications. For example, by controlling the chemical functionalization, dissipationless conduction ``wires'' could be patterned in an ultrathin tin film for low-power-consumption electronics.

First-principles calculations based on density functional theory were performed by the Vienna \emph{ab initio} simulation package~\cite{vasp-prb}, using the projector-augmented-wave potential with $4d$ electrons of tin described as valence and the plane-wave basis with an energy cutoff of 500 eV. The exchange-correlation functional was treated using the Perdew-Burke-Ernzerhof~\cite{perdew1996generalized} generalized-gradient approximation. The predicted topology was further verified by using the Heyd-Scuseria-Ernzerhof hybrid functional~\cite{krukau2006influence}. For tin films in QSH states, the two types of funcitonals predict essentially the same nontrivial bulk gaps, as these bulk gaps are opened by the spin-orbital coupling (SOC) effect. The SOC was included in the self-consistent calculations of electronic structure.

\begin{figure}
\includegraphics[width=\linewidth]{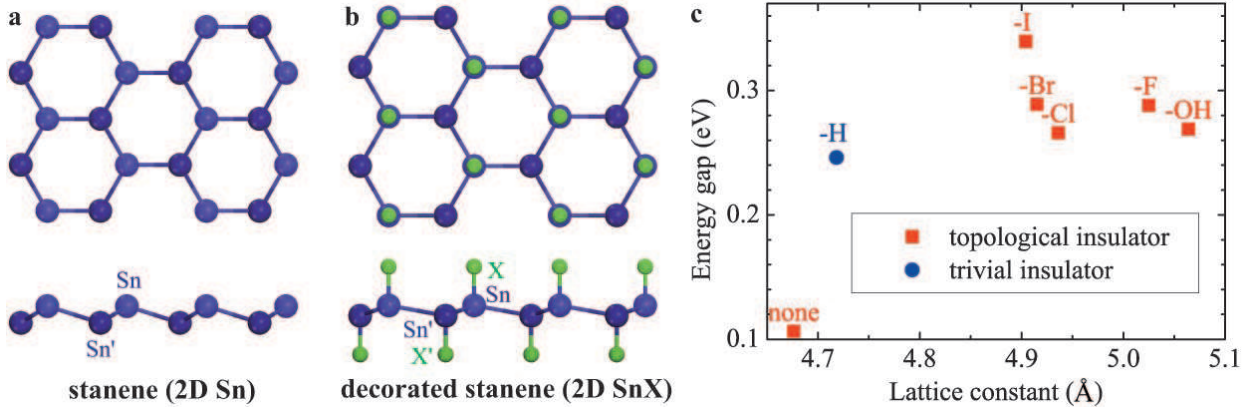}
\caption{(color online) (a,b) Crystal structure for stanene (2D Sn) and decorated stanene (2D SnX) from the top/side view (upper/lower).  X represents the chemical functional group. In a unit cell Sn (X) is related to Sn$'$ (X$'$) by an inversion operation. (c) The calculated energy gap for stanene (labeled by ``none'') and decorated stanene (labeled by ``-X'') at their equilibrium lattice constants. Red squares and blue circles mark topological and trivial insulators, respectively.}
\end{figure}

Figures 1a and 1b show optimal geometries for stanene and decorated stanene, denoted as 2D Sn and 2D SnX, respectively. X represents the chemical functional group. A low-buckled configuration is found to be more stable for stanene~\cite{cahangirov2009two}, in contrast to the planar geometry of graphene. This is related to the relatively weak $\pi$-$\pi$ bonding between tin atoms. The buckling enhances the overlap between $\pi$ and $\sigma$ orbitals and stabilizes the system. The same mechanism also applies for silicene that shares a similar configuration. For decorated stanene structures (like stanane), they have a stable $sp^3$ configuration analogous to graphane, with the chemical functional groups alternating on both sides of the nanosheet in their most stable configuration (see Fig. 1b). We further confirmed the stability of the 2D tin films by phonon calculations that show no imaginary frequency. Compared to stanane, in decorated stanene the Sn-Sn bond length slightly increases, the buckling of the tin nanosheet decreases, and the equilibrium lattice constant enhances as shown in Fig. 1c. Whereas the inversion symmetry holds for both types of systems.

\begin{figure}
\includegraphics[width=\linewidth]{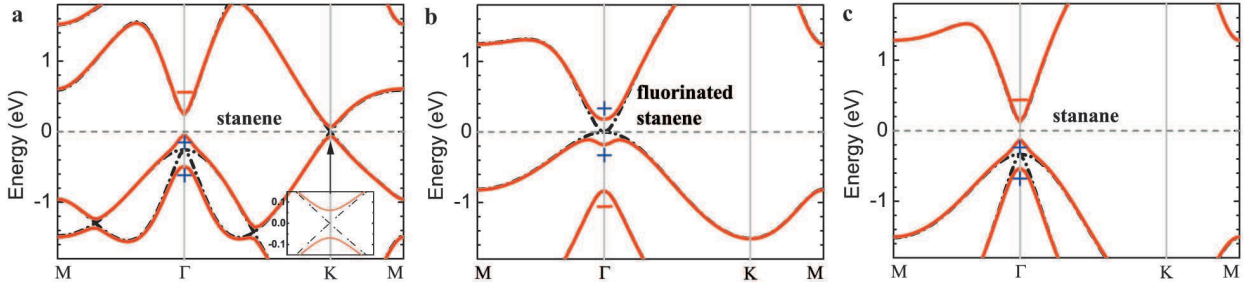}
\caption{(color online) Band structure for (a) stanene, (b) fluorinated stanene and (c) stanane without (black dash-dot lines) and with (red solid lines) spin-orbital coupling. The inset of (a) shows a zoomed in energy dispersion near the $\mathrm{K}$ point. The Fermi level is indicated by the dashed line. Parities of the Bloch states at the $\Gamma$ point are denoted by $+$,$-$.}
\end{figure}

The band structure of stanene is shown in Figure 2a. Two energy bands cross linearly at the $\mathbf{K}$ (and $\mathbf{K}' = -\mathbf{K}$) point without SOC; a band gap opens when the SOC is turned on. Thus stanene is a QSH insulator~\cite{liu2011low}, similar to graphene~\cite{kane2005quantum}. With the stronger SOC, stanene has a larger gap of 0.1 eV. Since the graphene analogue of lead (the heaviest element of group IV) is a metal, the value of 0.1 eV seems to be the largest nontrivial gap we could achieve for 2D group IV films. Such an upper limit, however, can be significantly broken through, as we will show.

Chemical functionalization of 2D materials is a powerful tool to create new materials with desirable features, as  graphane~\cite{elias2009control} or fluorinated graphene~\cite{robinson2010properties} to graphene.
The great flexibility in the selection of the chemical functional group enables forming a series of chemically new materials. We find the concept useful for designing new QSH insulators, particularly in 2D tin films. Compared to stanene, decorated (or functionalized) stanene structures offer much more possibilities. We considered some typical chemical functional groups for demonstration. As summarized in Fig. 1c, some chemical functional groups (e.g., X = -F, -Cl, -Br, -I and -OH) result in QSH insulators, whereas some others (e.g., X = -H) do not. Remarkably, the chemical functionalization creates QSH insulators with sizable nontrivial bulk gaps of 0.3 eV, considerably exceeding the presumed upper limit settled by the system without decoration.

To reveal the impressive effect of chemical functionalization, we take fluorinated stanene as an example. Figure 2b presents its band structure without and with SOC. Compared to stanene, one can clearly see that at the $\mathrm{K}$ point the band gap is substantially enlarged due to the saturation of $\pi$ orbital. Similar feature is observed for the fluorination of graphene. For graphene the chemical functionalization destroys the low-energy physics at the $\mathrm{K}$ point and drives the system into a trivial phase. The fluorinated graphene is a trivial insulator, whereas fluorinated stanene is not, which is confirmed by calculating the parities of the Bloch wavefunctions for occupied bands at all time-reversal invariant points (including one $\Gamma$ and three $\mathrm{M}$ points), based on the method proposed by Fu and Kane~\cite{fu2007topological_inversion}. Qualitative differences between the  two systems exist at the $\Gamma$ point. Fluorinated stanene is gapless with the valence and conduction bands degenerate at the $\Gamma$ point when excluding SOC, and becomes fully gapped when including SOC. In comparison, the band gap of the fluorinated graphene is large at the $\Gamma$ point and  negligibly affected by the SOC effect (data not shown). Moreover, for the 2D tin system the chemical functionalization induces a parity exchange between occupied and unoccupied bands at the $\Gamma$ point. As indicated in Figs. 2a and 2b, at the $\Gamma$ point a negative-parity Bloch state forms the conduction band minimum (CBM) for stanene and shifts downwards into valence bands upon the chemical functionalization, leaving a positive-parity Bloch state as the CBM for fluorinated stanene. Such a parity change, that is nontrivial to the topology~\cite{fu2007topological_inversion}, is not observed for the graphene system due to the large energy gap at the $\Gamma$ point.

\begin{figure}
\includegraphics[width=\linewidth]{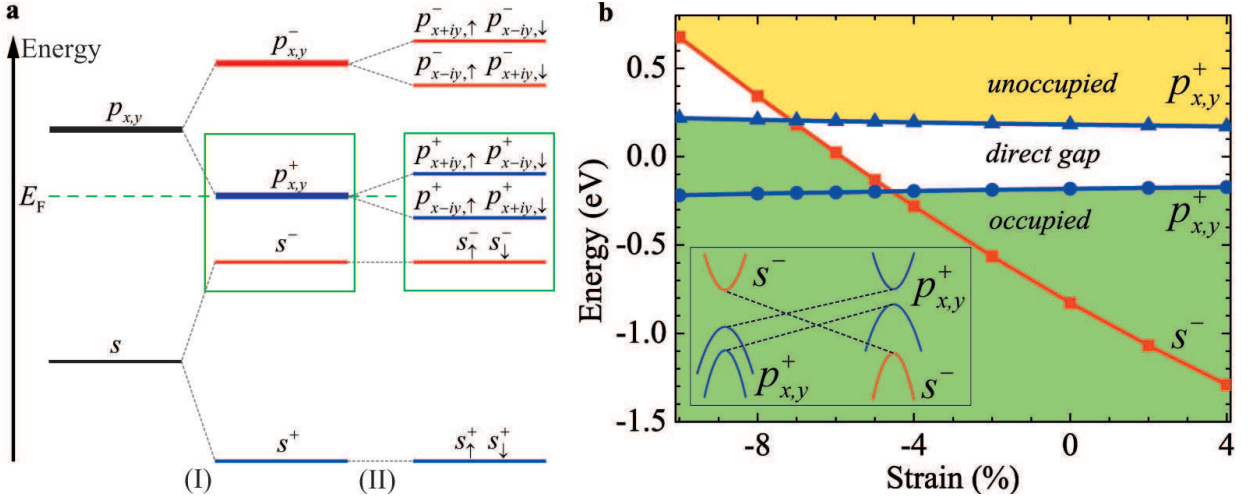}
\caption{(color online) (a) Schematic diagram of the evolution from the atomic $s$ and $p_{x,y}$ orbitals of Sn into the conduction and valence bands at the $\Gamma$ point for fluorinated stanene. The stages (I) and (II) represent the effect of turning on chemical bonding and spin-orbital coupling, respectively (see text). The green dashed line denotes the Fermi energy $E_\mathrm{F}$. (b) The energy levels $|s^{-}\rangle$ and $|p_{x,y}^{+}\rangle$ of fluorinated stanene at the $\Gamma$ point under different hydrostatic strains, with a schematic representation shown in the inset. The center of the two split $|p_{x,y}^{+}\rangle$ levels is defined as the energy zero.}
\end{figure}

To get a physical understanding on the topological nature, we start from atomic orbitals and consider the effect of chemical bonding and SOC on the energy levels at the $\Gamma$ point for fluorinated stanene. This is schematically illustrated in two stages (I) and (II) in Fig. 3a. States around the Fermi energy are mainly contributed by the atomic $s$ and $p_{x,y}$ orbitals of Sn ($5s^2 5p^2$), whose $p_z$ orbital is saturated by F ($2s^2 2p^5$). Therefore, we focus on $s$ and $p_{x,y}$ orbitals of Sn and neglect the effect of other atomic orbitals. In stage (I), the chemical bonding between Sn-Sn atoms forms bonding and antibonding states for both $s$ and $p_{x,y}$ orbitals.  These states are labeled as $|s^{\pm}\rangle$ and $|p_{x,y}^{\pm}\rangle$, where the superscript ($+$,$-$) denotes the parity. Energy levels close to the Fermi energy turn out to be $|s^{-}\rangle$ and $|p_{x,y}^{+}\rangle$. While $|s^{-}\rangle$ is typically above $|p_{x,y}^{+}\rangle$ in energy (like for graphane), the order is reversed in the present system. Before turning on SOC, the $p_x$ and $p_y$ orbitals is degenerate due to the $C_3$ rotation symmetry, and the system is a zero-gap semiconductor. As the SOC effect is taken into account in stage (II), the degeneracy of the $|p_{x,y}^{+}\rangle$ level is lifted, opening a full energy gap. This introduces a nontrivial insulating phase with an inverted order of $|s^{-}\rangle$ and $|p_{x,y}^{+}\rangle$.
The mechanism is similar as for HgTe quantum well~\cite{bernevig2006quantum}, Heusler compounds~\cite{chadov2010tunable,lin2010half} and KHgSb~\cite{yan2012prediction}, which also have the so-called s-p-type band inversion~\cite{zhang2013TI}.

To illustrate the band inversion process explicitly, we present in Fig. 3b the calculated energy levels $|s^{-}\rangle$ and $|p_{x,y}^{+}\rangle$ at the $\Gamma$ point for fluorinated stanene under different hydrostatic strains. When compressing the lattice, the antibonding state $|s^{-}\rangle$ shifts upwards with respect to the bonding state $|p_{x,y}^{+}\rangle$. A crossing between the $|s^{-}\rangle$ level and the upper $|p_{x,y}^{+}\rangle$ level occurs at a critical strain of around -7\%. This level crossing leads to a parity exchange between occupied and unoccupied bands, therefore, induces a topological phase transition from a TI to a trivial insulator.

As the interval between the energy levels $|s^{-}\rangle$ and $|p_{x,y}^{+}\rangle$ at the $\Gamma$ point sensitively depends on the bonding strength between Sn-Sn atoms, the band inversion and the associated QSH states can be tuned effectively by attaching different types of chemical functional groups and by applying external strain. At the equilibrium lattice constant, the band inversion exists for some chemical functional groups (like X = -F) whereas not for others (like X = -H, see Fig. 2c), resulting in topologically different phases as summarized in Fig. 1c. In addition, external strain can be introduced by deposition onto a substrate or by exerting mechanical forces to a suspended sample. Here the nontrivial gaps are opened between the split $|p_{x,y}^{+}\rangle$ states due to the SOC effect for Sn-Sn bonds. Their magnitude depends weakly on the type of chemical functional group, and keeps very large ($\sim$0.3 eV) owing to the strong SOC of tin.

\begin{figure}
\includegraphics[width=\linewidth]{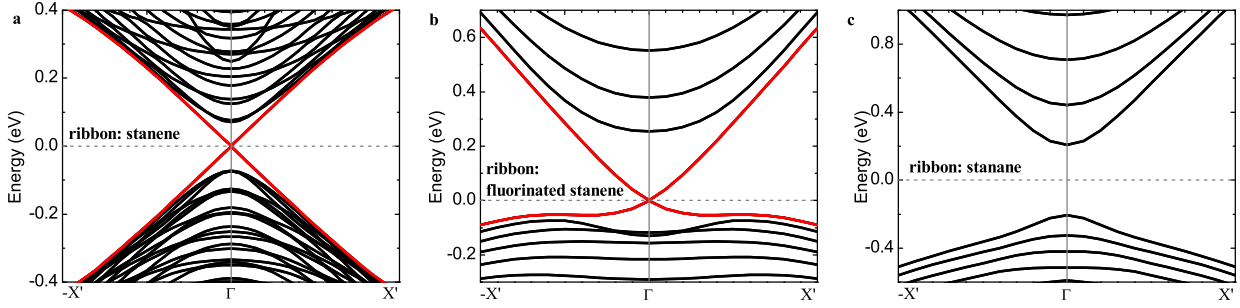}
\caption{(color online) Band structure of armchair-edge nanoribbons for (a) stanene, (b) fluorinated stanene and (c) stanane. The ribbon widths are 15.2, 8.4 and 7.8 nm, respectively. $\mathrm{X}' = 0.2 \pi/L_0$, where $L_0$ is the periodicity of the nanoribbon along its length direction. Helical edge states are visualized by red lines crossing linearly at the $\Gamma$ point for stanene and fluorinated stanene. No helical edge state exists for stanane.}
\end{figure}

One of prominent features of QSH insulators is the existence of gapless edge states that are helical with spin-momentum locked. To see the helical edge states explicitly, we construct nanoribbon structures with symmetric edges, for which edge states are degenerate due to the existence of two edges. The widths of nanoribbons are selected to be large enough to avoid interactions between edge states. Without loss of generality, we consider armchair-type edges, and saturate the dangling bonds of edge atoms by H for stanene and by the functional group ``-X'' for decorated stanene. The calculated band structures of nanoribbons are presented in Fig. 4. For stanene and fluorinated stanene, one can easily see helical edge states that form bands dispersing in the bulk gap and crossing linearly at the $\Gamma$ point. Each edge has a single pair of helical edge states for both systems. In comparison, no helical edge state exists for stanane. The results agree well with the parity analysis. Helical edge states are very useful for electronics and spintronics owing to their robustness against scattering. As an important quantity related to applications, the Fermi velocity of helical edge states  $v_\mathrm{F}$ is around $4.4 \times 10^5$ m/s for stanene and $6.8 \times 10^5$ m/s for fluorinated stanene. The latter value is significantly larger than that for HgTe quantum well ($5.5 \times 10^5$ m/s)~\cite{qi2011topological}, advantageous for  high-speed devices and circuits. Moreover, for the nontrivial decorated stanene systems $v_\mathrm{F}$ noticeably varies as modifying the chemical functional group . For instance, $v_\mathrm{F}$ varies from $\sim$5.2 to $\sim$7.2 $\times 10^5$ m/s if changing X = -I to X = -Cl.

The 2D tin films are useful for applications not only because of their large nontrivial bulk gaps, but also because of the tunability of QSH states by chemical functionalization. Varying the chemical functional group  quantitatively changes the Fermi velocity of helical edges states, or even qualitatively changes the topology, resulting in topologically different phases. On the experimental side, various techniques have been developed to grow 2D materials, including mechanical exfoliation, chemical exfoliation, molecule/atom intercalation, molecular beam epitaxy (MBE) and so on~\cite{Butler2013Progress}. Among them MBE is the most convenient way to grow the ultrathin tin materials, probably on substrates like hexagonal boron nitride, highly orientated pyrolytic graphite (HOPG), and the (111) surfaces of CdTe or InSb. Based on MBE the synthesis of silicene~\cite{vogt2012silicene} and possible formation of tin monolayer in graphene structure~\cite{osaka1994surface,zimmermann1997growth} were reported. Moreover, the chemical functionalization may be achieved by exposure to atomic or molecular gases or by chemical reaction in solvents.

Next we discuss the possible role of a substrate, as the tin film in experiment is usually supported by a substrate. The interactions with a substrate, when weak without forming any chemical bond (for instance on a monolayer hexagonal boron nitride), have little effect on the electronic structure (including the band gap and topological order) of a decorated
tin film. In contrast, the formation of chemical bonds with a substrate has a strong impact on the geometry and electronic structure of the tin film. In cases of the strong coupling, a decorated tin film keeps as a QSH insulator provided that its band structure has a band gap at the K point and also has a band inversion at the $\Gamma$ point. Once the Sn's $p_z$ orbitals are fully saturated either by forming chemical bonds with the substrate or by chemical functionalization, a band gap opens at the K point. Then the system is topologically nontrivial provided that a band inversion occurs at the $\Gamma$ point, which can be controlled effectively by strain as demonstrated in Fig. 3(b). The scenario is confirmed by our calculations for iodinated stanene (2D SnI) on the clean CdTe(111)B (Te-terminated) surface. In this strongly coupled system, the nontrivial topological phase remains observable, although the band gap reduces (to around 0.1 eV). Our results indicate that it is feasible to observe the nontrivial topological phase in the decorated tin film on a substrate.

\begin{figure}
\includegraphics[width=\linewidth]{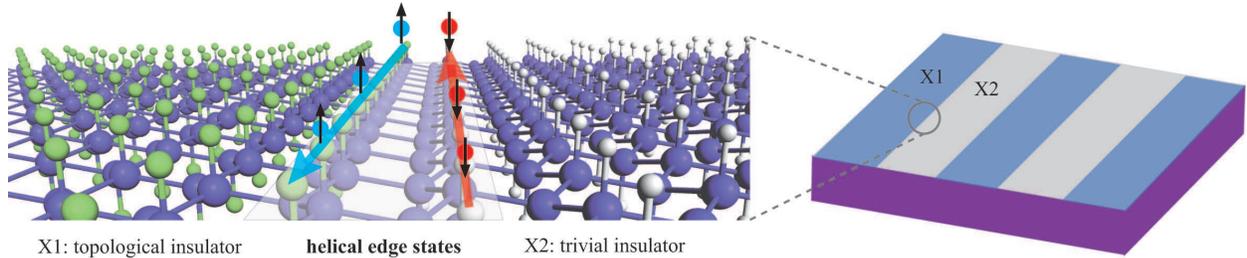}
\caption{(color online) A schematic diagram showing helical edge states at the phase boundary between topological insulator and trivial insulator, and are embedded in an atomically thin tin sheet with different chemical functional groups (e.g. X1 = -F and X2 = -H). The helical edge states can be patterned by controlling the chemical functionalization and used as dissipationless conducting ``wires'' for electronic circuits.}
\end{figure}

As one of potential applications, helical edge states, which appear at the phase boundary, can be embedded in stanene by using different chemical functional groups, as shown schematically in Fig. 5. Since the atomic-scale control over the chemical functionalization is feasible nowadays, helical edge states can be patterned in a controlled way to be used as dissipationless conducting ``wires'' for electronic circuits. The proposal, if realized, could significantly lower the power consumption and heat generation rates of electronic devices, which is crucial for the development of modern integrated circuits. Furthermore, breaking the time-reversal symmetry of the QSH state can lead to the quantum anomalous Hall (QAH) state~\cite{haldane1988model,qi2011topological,yu2010quantized,chang2013experimental} that is another novel quantum state supporting dissipationless conducting channels. The QAH state can be realized in tin films probably by surface doping of magnetic atoms or molecules, like Cr-doped Bi$_2$(Se$_x$Te$_{1-x}$)$_3$ TI films~\cite{zhang2013topology}, which will be discussed in future work.

This work is supported by the Defense Advanced Research Projects Agency Microsystems Technology Office, MesoDynamic Architecture Program (MESO) through the contract number N66001-11-1-4105 and by the DARPA Program on "Topological Insulators -- Solid State Chemistry, New Materials and Properties", under the award number N66001-12-1-4034.


\begin{thebibliography}{10}%
\makeatletter
\providecommand \@ifxundefined [1]{%
 \ifx #1\undefined \expandafter \@firstoftwo
 \else \expandafter \@secondoftwo
\fi
}%
\providecommand \@ifnum [1]{%
 \ifnum #1\expandafter \@firstoftwo
 \else \expandafter \@secondoftwo
\fi
}%
\providecommand \enquote [1]{``#1''}%
\providecommand \bibnamefont  [1]{#1}%
\providecommand \bibfnamefont [1]{#1}%
\providecommand \citenamefont [1]{#1}%
\providecommand\href[0]{\@sanitize\@href}%
\providecommand\@href[1]{\endgroup\@@startlink{#1}\endgroup\@@href}%
\providecommand\@@href[1]{#1\@@endlink}%
\providecommand \@sanitize [0]{\begingroup\catcode`\&12\catcode`\#12\relax}%
\@ifxundefined \pdfoutput {\@firstoftwo}{%
 \@ifnum{\z@=\pdfoutput}{\@firstoftwo}{\@secondoftwo}%
}{%
 \providecommand\@@startlink[1]{\leavevmode}%
 \providecommand\@@endlink[0]{}%
}{%
 \providecommand\@@startlink[1]{%
  \leavevmode
  \pdfstartlink
   attr{/Border[0 0 1 ]/H/I/C[0 1 1]}%
   user{/Subtype/Link/A<</Type/Action/S/URI/URI(#1)>>}%
  \relax
 }%
 \providecommand\@@endlink[0]{\pdfendlink}%
}%
\providecommand \url  [0]{\begingroup\@sanitize \@url }%
\providecommand \@url [1]{\endgroup\@href {#1}{\urlprefix}}%
\providecommand \urlprefix [0]{URL }%
\providecommand \Eprint[0]{\href }%
\@ifxundefined \urlstyle {%
  \providecommand \doi [1]{doi:\discretionary{}{}{}#1}%
}{%
  \providecommand \doi [0]{doi:\discretionary{}{}{}\begingroup
  \urlstyle{rm}\Url }%
}%
\providecommand \doibase [0]{http://dx.doi.org/}%
\providecommand \Doi[1]{\href{\doibase#1}}%
\providecommand \bibAnnote [3]{%
  \BibitemShut{#1}%
  \begin{quotation}\noindent
    \textsc{Key:}\ #2\\\textsc{Annotation:}\ #3%
  \end{quotation}%
}%
\providecommand \bibAnnoteFile [2]{%
  \IfFileExists{#2}{\bibAnnote {#1} {#2} {\input{#2}}}{}%
}%
\providecommand \typeout [0]{\immediate \write \m@ne }%
\providecommand \selectlanguage [0]{\@gobble}%
\providecommand \bibinfo [0]{\@secondoftwo}%
\providecommand \bibfield [0]{\@secondoftwo}%
\providecommand \translation [1]{[#1]}%
\providecommand \BibitemOpen[0]{}%
\providecommand \bibitemStop [0]{}%
\providecommand \bibitemNoStop [0]{.\EOS\space}%
\providecommand \EOS [0]{\spacefactor3000\relax}%
\providecommand \BibitemShut [1]{\csname bibitem#1\endcsname}%
\bibitem{qi2010quantum}%
  \BibitemOpen
  \bibfield{author}{%
  \bibinfo {author} {\bibfnamefont{X.-L.}\ \bibnamefont{Qi}}\ and\ \bibinfo
  {author} {\bibfnamefont{S.-C.}\ \bibnamefont{Zhang}},\ }%
  \bibfield{journal}{%
  \bibinfo {journal} {Phys. Today}\ }%
  \textbf{\bibinfo {volume} {63}},\ \bibinfo {pages} {33} (\bibinfo {year}
  {2010})%
  \bibAnnoteFile{NoStop}{qi2010quantum}%
\bibitem{hasan2010colloquium}%
  \BibitemOpen
  \bibfield{author}{%
  \bibinfo {author} {\bibfnamefont{M.~Z.}\ \bibnamefont{Hasan}}\ and\ \bibinfo
  {author} {\bibfnamefont{C.~L.}\ \bibnamefont{Kane}},\ }%
  \bibfield{journal}{%
  \bibinfo {journal} {Rev. Mod. Phys.}\ }%
  \textbf{\bibinfo {volume} {82}},\ \bibinfo {pages} {3045} (\bibinfo {year}
  {2010})%
  \bibAnnoteFile{NoStop}{hasan2010colloquium}%
\bibitem{qi2011topological}%
  \BibitemOpen
  \bibfield{author}{%
  \bibinfo {author} {\bibfnamefont{X.-L.}\ \bibnamefont{Qi}}\ and\ \bibinfo
  {author} {\bibfnamefont{S.-C.}\ \bibnamefont{Zhang}},\ }%
  \bibfield{journal}{%
  \bibinfo {journal} {Rev. Mod. Phys.}\ }%
  \textbf{\bibinfo {volume} {83}},\ \bibinfo {pages} {1057} (\bibinfo {year}
  {2011})%
  \bibAnnoteFile{NoStop}{qi2011topological}%
\bibitem{yan2012topological}%
  \BibitemOpen
  \bibfield{author}{%
  \bibinfo {author} {\bibfnamefont{B.}~\bibnamefont{Yan}}\ and\ \bibinfo
  {author} {\bibfnamefont{S.-C.}\ \bibnamefont{Zhang}},\ }%
  \bibfield{journal}{%
  \bibinfo {journal} {Rep. Prog. Phys.}\ }%
  \textbf{\bibinfo {volume} {75}},\ \bibinfo {pages} {96501} (\bibinfo {year}
  {2012})%
  \bibAnnoteFile{NoStop}{yan2012topological}%
\bibitem{bernevig2006quantum}%
  \BibitemOpen
  \bibfield{author}{%
  \bibinfo {author} {\bibfnamefont{B.~A.}\ \bibnamefont{Bernevig}}, \bibinfo
  {author} {\bibfnamefont{T.~L.}\ \bibnamefont{Hughes}},\ and\ \bibinfo
  {author} {\bibfnamefont{S.-C.}\ \bibnamefont{Zhang}},\ }%
  \bibfield{journal}{%
  \bibinfo {journal} {Science}\ }%
  \textbf{\bibinfo {volume} {314}},\ \bibinfo {pages} {1757} (\bibinfo {year}
  {2006})%
  \bibAnnoteFile{NoStop}{bernevig2006quantum}%
\bibitem{konig2007quantum}%
  \BibitemOpen
  \bibfield{author}{%
  \bibinfo {author} {\bibfnamefont{M.}~\bibnamefont{K{\"o}nig}}, \bibinfo
  {author} {\bibfnamefont{S.}~\bibnamefont{Wiedmann}}, \bibinfo {author}
  {\bibfnamefont{C.}~\bibnamefont{Br{\"u}ne}}, \bibinfo {author}
  {\bibfnamefont{A.}~\bibnamefont{Roth}}, \bibinfo {author}
  {\bibfnamefont{H.}~\bibnamefont{Buhmann}}, \bibinfo {author}
  {\bibfnamefont{L.~W.}\ \bibnamefont{Molenkamp}}, \bibinfo {author}
  {\bibfnamefont{X.-L.}\ \bibnamefont{Qi}},\ and\ \bibinfo {author}
  {\bibfnamefont{S.-C.}\ \bibnamefont{Zhang}},\ }%
  \bibfield{journal}{%
  \bibinfo {journal} {Science}\ }%
  \textbf{\bibinfo {volume} {318}},\ \bibinfo {pages} {766} (\bibinfo {year}
  {2007})%
  \bibAnnoteFile{NoStop}{konig2007quantum}%
\bibitem{Bi2X3-natphy}%
  \BibitemOpen
  \bibfield{author}{%
  \bibinfo {author} {\bibfnamefont{H.}~\bibnamefont{Zhang}}, \bibinfo {author}
  {\bibfnamefont{C.-X.}\ \bibnamefont{Liu}}, \bibinfo {author}
  {\bibfnamefont{X.-L.}\ \bibnamefont{Qi}}, \bibinfo {author}
  {\bibfnamefont{X.}~\bibnamefont{Dai}}, \bibinfo {author}
  {\bibfnamefont{Z.}~\bibnamefont{Fang}},\ and\ \bibinfo {author}
  {\bibfnamefont{S.-C.}\ \bibnamefont{Zhang}},\ }%
  \bibfield{journal}{%
  \bibinfo {journal} {Nature Phys.}\ }%
  \textbf{\bibinfo {volume} {5}},\ \bibinfo {pages} {438} (\bibinfo {year}
  {2009})%
  \bibAnnoteFile{NoStop}{Bi2X3-natphy}%
\bibitem{Bi2Te3-exp-Chen}%
  \BibitemOpen
  \bibfield{author}{%
  \bibinfo {author} {\bibfnamefont{Y.~L.}\ \bibnamefont{Chen}}, \bibinfo
  {author} {\bibfnamefont{J.~G.}\ \bibnamefont{Analytis}}, \bibinfo {author}
  {\bibfnamefont{J.-H.}\ \bibnamefont{Chu}}, \bibinfo {author}
  {\bibfnamefont{Z.~K.}\ \bibnamefont{Liu}}, \bibinfo {author}
  {\bibfnamefont{S.-K.}\ \bibnamefont{Mo}}, \bibinfo {author}
  {\bibfnamefont{X.~L.}\ \bibnamefont{Qi}}, \bibinfo {author}
  {\bibfnamefont{H.~J.}\ \bibnamefont{Zhang}}, \bibinfo {author}
  {\bibfnamefont{D.~H.}\ \bibnamefont{Lu}}, \bibinfo {author}
  {\bibfnamefont{X.}~\bibnamefont{Dai}}, \bibinfo {author}
  {\bibfnamefont{Z.}~\bibnamefont{Fang}}, \bibinfo {author}
  {\bibfnamefont{S.~C.}\ \bibnamefont{Zhang}}, \bibinfo {author}
  {\bibfnamefont{I.~R.}\ \bibnamefont{Fisher}}, \bibinfo {author}
  {\bibfnamefont{Z.}~\bibnamefont{Hussain}},\ and\ \bibinfo {author}
  {\bibfnamefont{Z.-X.}\ \bibnamefont{Shen}},\ }%
  \bibfield{journal}{%
  \bibinfo {journal} {Science}\ }%
  \textbf{\bibinfo {volume} {325}},\ \bibinfo {pages} {178} (\bibinfo {year}
  {2009})%
  \bibAnnoteFile{NoStop}{Bi2Te3-exp-Chen}%
\bibitem{Bi2Se3-exp-Hasan}%
  \BibitemOpen
  \bibfield{author}{%
  \bibinfo {author} {\bibfnamefont{Y.}~\bibnamefont{Xia}}, \bibinfo {author}
  {\bibfnamefont{D.}~\bibnamefont{Qian}}, \bibinfo {author}
  {\bibfnamefont{D.}~\bibnamefont{Hsieh}}, \bibinfo {author}
  {\bibfnamefont{L.}~\bibnamefont{Wray}}, \bibinfo {author}
  {\bibfnamefont{A.}~\bibnamefont{Pal}}, \bibinfo {author}
  {\bibfnamefont{H.}~\bibnamefont{Lin}}, \bibinfo {author}
  {\bibfnamefont{A.}~\bibnamefont{Bansil}}, \bibinfo {author}
  {\bibfnamefont{D.}~\bibnamefont{Grauer}}, \bibinfo {author}
  {\bibfnamefont{Y.~S.}\ \bibnamefont{Hor}}, \bibinfo {author}
  {\bibfnamefont{R.~J.}\ \bibnamefont{Cava}},\ and\ \bibinfo {author}
  {\bibfnamefont{M.~Z.}\ \bibnamefont{Hasan}},\ }%
  \bibfield{journal}{%
  \bibinfo {journal} {Nature Phys.}\ }%
  \textbf{\bibinfo {volume} {5}},\ \bibinfo {pages} {398} (\bibinfo {year}
  {2009})%
  \bibAnnoteFile{NoStop}{Bi2Se3-exp-Hasan}%
\bibitem{kane2005quantum}%
  \BibitemOpen
  \bibfield{author}{%
  \bibinfo {author} {\bibfnamefont{C.~L.}\ \bibnamefont{Kane}}\ and\ \bibinfo
  {author} {\bibfnamefont{E.~J.}\ \bibnamefont{Mele}},\ }%
  \bibfield{journal}{%
  \bibinfo {journal} {Phys. Rev. Lett.}\ }%
  \textbf{\bibinfo {volume} {95}},\ \bibinfo {pages} {226801} (\bibinfo {year}
  {2005})%
  \bibAnnoteFile{NoStop}{kane2005quantum}%
\bibitem{murakami2006quantum}%
  \BibitemOpen
  \bibfield{author}{%
  \bibinfo {author} {\bibfnamefont{S.}~\bibnamefont{Murakami}},\ }%
  \bibfield{journal}{%
  \bibinfo {journal} {Phys. Rev. Lett.}\ }%
  \textbf{\bibinfo {volume} {97}},\ \bibinfo {pages} {236805} (\bibinfo {year}
  {2006})%
  \bibAnnoteFile{NoStop}{murakami2006quantum}%
\bibitem{liu2011stable}%
  \BibitemOpen
  \bibfield{author}{%
  \bibinfo {author} {\bibfnamefont{Z.}~\bibnamefont{Liu}}, \bibinfo {author}
  {\bibfnamefont{C.-X.}\ \bibnamefont{Liu}}, \bibinfo {author}
  {\bibfnamefont{Y.-S.}\ \bibnamefont{Wu}}, \bibinfo {author}
  {\bibfnamefont{W.-H.}\ \bibnamefont{Duan}}, \bibinfo {author}
  {\bibfnamefont{F.}~\bibnamefont{Liu}},\ and\ \bibinfo {author}
  {\bibfnamefont{J.}~\bibnamefont{Wu}},\ }%
  \bibfield{journal}{%
  \bibinfo {journal} {Phys. Rev. Lett.}\ }%
  \textbf{\bibinfo {volume} {107}},\ \bibinfo {pages} {136805} (\bibinfo {year}
  {2011})%
  \bibAnnoteFile{NoStop}{liu2011stable}%
\bibitem{liu2008quantum}%
  \BibitemOpen
  \bibfield{author}{%
  \bibinfo {author} {\bibfnamefont{C.}~\bibnamefont{Liu}}, \bibinfo {author}
  {\bibfnamefont{T.~L.}\ \bibnamefont{Hughes}}, \bibinfo {author}
  {\bibfnamefont{X.-L.}\ \bibnamefont{Qi}}, \bibinfo {author}
  {\bibfnamefont{K.}~\bibnamefont{Wang}},\ and\ \bibinfo {author}
  {\bibfnamefont{S.-C.}\ \bibnamefont{Zhang}},\ }%
  \bibfield{journal}{%
  \bibinfo {journal} {Phys. Rev. Lett.}\ }%
  \textbf{\bibinfo {volume} {100}},\ \bibinfo {pages} {236601} (\bibinfo {year}
  {2008})%
  \bibAnnoteFile{NoStop}{liu2008quantum}%
\bibitem{knez2011evidence}%
  \BibitemOpen
  \bibfield{author}{%
  \bibinfo {author} {\bibfnamefont{I.}~\bibnamefont{Knez}}, \bibinfo {author}
  {\bibfnamefont{R.-R.}\ \bibnamefont{Du}},\ and\ \bibinfo {author}
  {\bibfnamefont{G.}~\bibnamefont{Sullivan}},\ }%
  \bibfield{journal}{%
  \bibinfo {journal} {Physical review letters}\ }%
  \textbf{\bibinfo {volume} {107}},\ \bibinfo {pages} {136603} (\bibinfo {year}
  {2011})%
  \bibAnnoteFile{NoStop}{knez2011evidence}%
\bibitem{xiao2011interface}%
  \BibitemOpen
  \bibfield{author}{%
  \bibinfo {author} {\bibfnamefont{D.}~\bibnamefont{Xiao}}, \bibinfo {author}
  {\bibfnamefont{W.}~\bibnamefont{Zhu}}, \bibinfo {author}
  {\bibfnamefont{Y.}~\bibnamefont{Ran}}, \bibinfo {author}
  {\bibfnamefont{N.}~\bibnamefont{Nagaosa}},\ and\ \bibinfo {author}
  {\bibfnamefont{S.}~\bibnamefont{Okamoto}},\ }%
  \bibfield{journal}{%
  \bibinfo {journal} {Nat. Commun.}\ }%
  \textbf{\bibinfo {volume} {2}},\ \bibinfo {pages} {596} (\bibinfo {year}
  {2011})%
  \bibAnnoteFile{NoStop}{xiao2011interface}%
\bibitem{liu2011quantum}%
  \BibitemOpen
  \bibfield{author}{%
  \bibinfo {author} {\bibfnamefont{C.-C.}\ \bibnamefont{Liu}}, \bibinfo
  {author} {\bibfnamefont{W.}~\bibnamefont{Feng}},\ and\ \bibinfo {author}
  {\bibfnamefont{Y.}~\bibnamefont{Yao}},\ }%
  \bibfield{journal}{%
  \bibinfo {journal} {Phys. Rev. Lett.}\ }%
  \textbf{\bibinfo {volume} {107}},\ \bibinfo {pages} {076802} (\bibinfo {year}
  {2011})%
  \bibAnnoteFile{NoStop}{liu2011quantum}%
\bibitem{novoselov2005two}%
  \BibitemOpen
  \bibfield{author}{%
  \bibinfo {author} {\bibfnamefont{K.}~\bibnamefont{Novoselov}}, \bibinfo
  {author} {\bibfnamefont{D.}~\bibnamefont{Jiang}}, \bibinfo {author}
  {\bibfnamefont{F.}~\bibnamefont{Schedin}}, \bibinfo {author}
  {\bibfnamefont{T.}~\bibnamefont{Booth}}, \bibinfo {author}
  {\bibfnamefont{V.}~\bibnamefont{Khotkevich}}, \bibinfo {author}
  {\bibfnamefont{S.}~\bibnamefont{Morozov}},\ and\ \bibinfo {author}
  {\bibfnamefont{A.}~\bibnamefont{Geim}},\ }%
  \bibfield{journal}{%
  \bibinfo {journal} {Proc. Natl. Acad. Sci.}\ }%
  \textbf{\bibinfo {volume} {102}},\ \bibinfo {pages} {10451} (\bibinfo {year}
  {2005})%
  \bibAnnoteFile{NoStop}{novoselov2005two}%
\bibitem{Butler2013Progress}%
  \BibitemOpen
  \bibfield{author}{%
  \bibinfo {author} {\bibfnamefont{S.~Z.}\ \bibnamefont{Butler}}, \bibinfo
  {author} {\bibfnamefont{S.~M.}\ \bibnamefont{Hollen}}, \bibinfo {author}
  {\bibfnamefont{L.}~\bibnamefont{Cao}}, \bibinfo {author}
  {\bibfnamefont{Y.}~\bibnamefont{Cui}}, \bibinfo {author}
  {\bibfnamefont{J.~A.}\ \bibnamefont{Gupta}}, \bibinfo {author}
  {\bibfnamefont{H.~R.}\ \bibnamefont{Guti¨¦rrez}}, \bibinfo {author}
  {\bibfnamefont{T.~F.}\ \bibnamefont{Heinz}}, \bibinfo {author}
  {\bibfnamefont{S.~S.}\ \bibnamefont{Hong}}, \bibinfo {author}
  {\bibfnamefont{J.}~\bibnamefont{Huang}}, \bibinfo {author}
  {\bibfnamefont{A.~F.}\ \bibnamefont{Ismach}}, \bibinfo {author}
  {\bibfnamefont{E.}~\bibnamefont{Johnston-Halperin}}, \bibinfo {author}
  {\bibfnamefont{M.}~\bibnamefont{Kuno}}, \bibinfo {author}
  {\bibfnamefont{V.~V.}\ \bibnamefont{Plashnitsa}}, \bibinfo {author}
  {\bibfnamefont{R.~D.}\ \bibnamefont{Robinson}}, \bibinfo {author}
  {\bibfnamefont{R.~S.}\ \bibnamefont{Ruoff}}, \bibinfo {author}
  {\bibfnamefont{S.}~\bibnamefont{Salahuddin}}, \bibinfo {author}
  {\bibfnamefont{J.}~\bibnamefont{Shan}}, \bibinfo {author}
  {\bibfnamefont{L.}~\bibnamefont{Shi}}, \bibinfo {author}
  {\bibfnamefont{M.~G.}\ \bibnamefont{Spencer}}, \bibinfo {author}
  {\bibfnamefont{M.}~\bibnamefont{Terrones}}, \bibinfo {author}
  {\bibfnamefont{W.}~\bibnamefont{Windl}},\ and\ \bibinfo {author}
  {\bibfnamefont{J.~E.}\ \bibnamefont{Goldberger}},\ }%
  \bibfield{journal}{%
  \Doi{10.1021/nn400280c}{\bibinfo {journal} {ACS Nano}}\ }%
  \textbf{\bibinfo {volume} {7}},\ \bibinfo {pages} {2898} (\bibinfo {year}
  {2013})%
  \bibAnnoteFile{NoStop}{Butler2013Progress}%
\bibitem{cahangirov2009two}%
  \BibitemOpen
  \bibfield{author}{%
  \bibinfo {author} {\bibfnamefont{S.}~\bibnamefont{Cahangirov}}, \bibinfo
  {author} {\bibfnamefont{M.}~\bibnamefont{Topsakal}}, \bibinfo {author}
  {\bibfnamefont{E.}~\bibnamefont{Akt{\"u}rk}}, \bibinfo {author}
  {\bibfnamefont{H.}~\bibnamefont{{\c{S}}ahin}},\ and\ \bibinfo {author}
  {\bibfnamefont{S.}~\bibnamefont{Ciraci}},\ }%
  \bibfield{journal}{%
  \bibinfo {journal} {Phys. Rev. Lett.}\ }%
  \textbf{\bibinfo {volume} {102}},\ \bibinfo {pages} {236804} (\bibinfo {year}
  {2009})%
  \bibAnnoteFile{NoStop}{cahangirov2009two}%
\bibitem{elias2009control}%
  \BibitemOpen
  \bibfield{author}{%
  \bibinfo {author} {\bibfnamefont{D.}~\bibnamefont{Elias}}, \bibinfo {author}
  {\bibfnamefont{R.}~\bibnamefont{Nair}}, \bibinfo {author}
  {\bibfnamefont{T.}~\bibnamefont{Mohiuddin}}, \bibinfo {author}
  {\bibfnamefont{S.}~\bibnamefont{Morozov}}, \bibinfo {author}
  {\bibfnamefont{P.}~\bibnamefont{Blake}}, \bibinfo {author}
  {\bibfnamefont{M.}~\bibnamefont{Halsall}}, \bibinfo {author}
  {\bibfnamefont{A.}~\bibnamefont{Ferrari}}, \bibinfo {author}
  {\bibfnamefont{D.}~\bibnamefont{Boukhvalov}}, \bibinfo {author}
  {\bibfnamefont{M.}~\bibnamefont{Katsnelson}}, \bibinfo {author}
  {\bibfnamefont{A.}~\bibnamefont{Geim}}, \emph{et~al.},\ }%
  \bibfield{journal}{%
  \bibinfo {journal} {Science}\ }%
  \textbf{\bibinfo {volume} {323}},\ \bibinfo {pages} {610} (\bibinfo {year}
  {2009})%
  \bibAnnoteFile{NoStop}{elias2009control}%
\bibitem{vogt2012silicene}%
  \BibitemOpen
  \bibfield{author}{%
  \bibinfo {author} {\bibfnamefont{P.}~\bibnamefont{Vogt}}, \bibinfo {author}
  {\bibfnamefont{P.}~\bibnamefont{De~Padova}}, \bibinfo {author}
  {\bibfnamefont{C.}~\bibnamefont{Quaresima}}, \bibinfo {author}
  {\bibfnamefont{J.}~\bibnamefont{Avila}}, \bibinfo {author}
  {\bibfnamefont{E.}~\bibnamefont{Frantzeskakis}}, \bibinfo {author}
  {\bibfnamefont{M.~C.}\ \bibnamefont{Asensio}}, \bibinfo {author}
  {\bibfnamefont{A.}~\bibnamefont{Resta}}, \bibinfo {author}
  {\bibfnamefont{B.}~\bibnamefont{Ealet}},\ and\ \bibinfo {author}
  {\bibfnamefont{G.}~\bibnamefont{Le~Lay}},\ }%
  \bibfield{journal}{%
  \bibinfo {journal} {Phys. Rev. Lett.}\ }%
  \textbf{\bibinfo {volume} {108}},\ \bibinfo {pages} {155501} (\bibinfo {year}
  {2012})%
  \bibAnnoteFile{NoStop}{vogt2012silicene}%
\bibitem{bianco2013stability}%
  \BibitemOpen
  \bibfield{author}{%
  \bibinfo {author} {\bibfnamefont{E.}~\bibnamefont{Bianco}}, \bibinfo {author}
  {\bibfnamefont{S.}~\bibnamefont{Butler}}, \bibinfo {author}
  {\bibfnamefont{S.}~\bibnamefont{Jiang}}, \bibinfo {author}
  {\bibfnamefont{O.~D.}\ \bibnamefont{Restrepo}}, \bibinfo {author}
  {\bibfnamefont{W.}~\bibnamefont{Windl}},\ and\ \bibinfo {author}
  {\bibfnamefont{J.~E.}\ \bibnamefont{Goldberger}},\ }%
  \bibfield{journal}{%
  \bibinfo {journal} {ACS Nano}\ }%
  \textbf{\bibinfo {volume} {7}},\ \bibinfo {pages} {4414} (\bibinfo {year}
  {2013})%
  \bibAnnoteFile{NoStop}{bianco2013stability}%
\bibitem{osaka1994surface}%
  \BibitemOpen
  \bibfield{author}{%
  \bibinfo {author} {\bibfnamefont{T.}~\bibnamefont{Osaka}}, \bibinfo {author}
  {\bibfnamefont{H.}~\bibnamefont{Omi}}, \bibinfo {author}
  {\bibfnamefont{K.}~\bibnamefont{Yamamoto}},\ and\ \bibinfo {author}
  {\bibfnamefont{A.}~\bibnamefont{Ohtake}},\ }%
  \bibfield{journal}{%
  \bibinfo {journal} {Phys. Rev. B}\ }%
  \textbf{\bibinfo {volume} {50}},\ \bibinfo {pages} {7567} (\bibinfo {year}
  {1994})%
  \bibAnnoteFile{NoStop}{osaka1994surface}%
\bibitem{zimmermann1997growth}%
  \BibitemOpen
  \bibfield{author}{%
  \bibinfo {author} {\bibfnamefont{H.}~\bibnamefont{Zimmermann}}, \bibinfo
  {author} {\bibfnamefont{R.~C.}\ \bibnamefont{Keller}}, \bibinfo {author}
  {\bibfnamefont{P.}~\bibnamefont{Meisen}},\ and\ \bibinfo {author}
  {\bibfnamefont{M.}~\bibnamefont{Seelmann-Eggebert}},\ }%
  \bibfield{journal}{%
  \bibinfo {journal} {Surf. Sci.}\ }%
  \textbf{\bibinfo {volume} {377}},\ \bibinfo {pages} {904} (\bibinfo {year}
  {1997})%
  \bibAnnoteFile{NoStop}{zimmermann1997growth}%
\bibitem{vasp-prb}%
  \BibitemOpen
  \bibfield{author}{%
  \bibinfo {author} {\bibfnamefont{G.}~\bibnamefont{Kresse}}\ and\ \bibinfo
  {author} {\bibfnamefont{J.}~\bibnamefont{Furthmuller}},\ }%
  \bibfield{journal}{%
  \bibinfo {journal} {Phys. Rev. B}\ }%
  \textbf{\bibinfo {volume} {54}},\ \bibinfo {pages} {11169} (\bibinfo {year}
  {1996})%
  \bibAnnoteFile{NoStop}{vasp-prb}%
\bibitem{perdew1996generalized}%
  \BibitemOpen
  \bibfield{author}{%
  \bibinfo {author} {\bibfnamefont{J.~P.}\ \bibnamefont{Perdew}}, \bibinfo
  {author} {\bibfnamefont{K.}~\bibnamefont{Burke}},\ and\ \bibinfo {author}
  {\bibfnamefont{M.}~\bibnamefont{Ernzerhof}},\ }%
  \bibfield{journal}{%
  \bibinfo {journal} {Phys. Rev. Lett.}\ }%
  \textbf{\bibinfo {volume} {77}},\ \bibinfo {pages} {3865} (\bibinfo {year}
  {1996})%
  \bibAnnoteFile{NoStop}{perdew1996generalized}%
\bibitem{krukau2006influence}%
  \BibitemOpen
  \bibfield{author}{%
  \bibinfo {author} {\bibfnamefont{A.~V.}\ \bibnamefont{Krukau}}, \bibinfo
  {author} {\bibfnamefont{O.~A.}\ \bibnamefont{Vydrov}}, \bibinfo {author}
  {\bibfnamefont{A.~F.}\ \bibnamefont{Izmaylov}},\ and\ \bibinfo {author}
  {\bibfnamefont{G.~E.}\ \bibnamefont{Scuseria}},\ }%
  \bibfield{journal}{%
  \bibinfo {journal} {J. Chem. Phys.}\ }%
  \textbf{\bibinfo {volume} {125}},\ \bibinfo {pages} {224106} (\bibinfo {year}
  {2006})%
  \bibAnnoteFile{NoStop}{krukau2006influence}%
\bibitem{liu2011low}%
  \BibitemOpen
  \bibfield{author}{%
  \bibinfo {author} {\bibfnamefont{C.-C.}\ \bibnamefont{Liu}}, \bibinfo
  {author} {\bibfnamefont{H.}~\bibnamefont{Jiang}},\ and\ \bibinfo {author}
  {\bibfnamefont{Y.}~\bibnamefont{Yao}},\ }%
  \bibfield{journal}{%
  \bibinfo {journal} {Phys. Rev. B}\ }%
  \textbf{\bibinfo {volume} {84}},\ \bibinfo {pages} {195430} (\bibinfo {year}
  {2011})%
  \bibAnnoteFile{NoStop}{liu2011low}%
\bibitem{robinson2010properties}%
  \BibitemOpen
  \bibfield{author}{%
  \bibinfo {author} {\bibfnamefont{J.~T.}\ \bibnamefont{Robinson}}, \bibinfo
  {author} {\bibfnamefont{J.~S.}\ \bibnamefont{Burgess}}, \bibinfo {author}
  {\bibfnamefont{C.~E.}\ \bibnamefont{Junkermeier}}, \bibinfo {author}
  {\bibfnamefont{S.~C.}\ \bibnamefont{Badescu}}, \bibinfo {author}
  {\bibfnamefont{T.~L.}\ \bibnamefont{Reinecke}}, \bibinfo {author}
  {\bibfnamefont{F.~K.}\ \bibnamefont{Perkins}}, \bibinfo {author}
  {\bibfnamefont{M.~K.}\ \bibnamefont{Zalalutdniov}}, \bibinfo {author}
  {\bibfnamefont{J.~W.}\ \bibnamefont{Baldwin}}, \bibinfo {author}
  {\bibfnamefont{J.~C.}\ \bibnamefont{Culbertson}}, \bibinfo {author}
  {\bibfnamefont{P.~E.}\ \bibnamefont{Sheehan}}, \emph{et~al.},\ }%
  \bibfield{journal}{%
  \bibinfo {journal} {Nano lett.}\ }%
  \textbf{\bibinfo {volume} {10}},\ \bibinfo {pages} {3001} (\bibinfo {year}
  {2010})%
  \bibAnnoteFile{NoStop}{robinson2010properties}%
\bibitem{fu2007topological_inversion}%
  \BibitemOpen
  \bibfield{author}{%
  \bibinfo {author} {\bibfnamefont{L.}~\bibnamefont{Fu}}\ and\ \bibinfo
  {author} {\bibfnamefont{C.~L.}\ \bibnamefont{Kane}},\ }%
  \bibfield{journal}{%
  \bibinfo {journal} {Phys. Rev. B}\ }%
  \textbf{\bibinfo {volume} {76}},\ \bibinfo {pages} {045302} (\bibinfo {year}
  {2007})%
  \bibAnnoteFile{NoStop}{fu2007topological_inversion}%
\bibitem{chadov2010tunable}%
  \BibitemOpen
  \bibfield{author}{%
  \bibinfo {author} {\bibfnamefont{S.}~\bibnamefont{Chadov}}, \bibinfo {author}
  {\bibfnamefont{X.}~\bibnamefont{Qi}}, \bibinfo {author}
  {\bibfnamefont{J.}~\bibnamefont{K{\"u}bler}}, \bibinfo {author}
  {\bibfnamefont{G.~H.}\ \bibnamefont{Fecher}}, \bibinfo {author}
  {\bibfnamefont{C.}~\bibnamefont{Felser}},\ and\ \bibinfo {author}
  {\bibfnamefont{S.~C.}\ \bibnamefont{Zhang}},\ }%
  \bibfield{journal}{%
  \bibinfo {journal} {Nature Mater.}\ }%
  \textbf{\bibinfo {volume} {9}},\ \bibinfo {pages} {541} (\bibinfo {year}
  {2010})%
  \bibAnnoteFile{NoStop}{chadov2010tunable}%
\bibitem{lin2010half}%
  \BibitemOpen
  \bibfield{author}{%
  \bibinfo {author} {\bibfnamefont{H.}~\bibnamefont{Lin}}, \bibinfo {author}
  {\bibfnamefont{L.~A.}\ \bibnamefont{Wray}}, \bibinfo {author}
  {\bibfnamefont{Y.}~\bibnamefont{Xia}}, \bibinfo {author}
  {\bibfnamefont{S.}~\bibnamefont{Xu}}, \bibinfo {author}
  {\bibfnamefont{S.}~\bibnamefont{Jia}}, \bibinfo {author}
  {\bibfnamefont{R.~J.}\ \bibnamefont{Cava}}, \bibinfo {author}
  {\bibfnamefont{A.}~\bibnamefont{Bansil}},\ and\ \bibinfo {author}
  {\bibfnamefont{M.~Z.}\ \bibnamefont{Hasan}},\ }%
  \bibfield{journal}{%
  \bibinfo {journal} {Nature Mater.}\ }%
  \textbf{\bibinfo {volume} {9}},\ \bibinfo {pages} {546} (\bibinfo {year}
  {2010})%
  \bibAnnoteFile{NoStop}{lin2010half}%
\bibitem{yan2012prediction}%
  \BibitemOpen
  \bibfield{author}{%
  \bibinfo {author} {\bibfnamefont{B.}~\bibnamefont{Yan}}, \bibinfo {author}
  {\bibfnamefont{L.}~\bibnamefont{M\"uchler}},\ and\ \bibinfo {author}
  {\bibfnamefont{C.}~\bibnamefont{Felser}},\ }%
  \bibfield{journal}{%
  \bibinfo {journal} {Phys. Rev. Lett.}\ }%
  \textbf{\bibinfo {volume} {109}},\ \bibinfo {pages} {116406} (\bibinfo {year}
  {2012})%
  \bibAnnoteFile{NoStop}{yan2012prediction}%
\bibitem{zhang2013TI}%
  \BibitemOpen
  \bibfield{author}{%
  \bibinfo {author} {\bibfnamefont{H.}~\bibnamefont{Zhang}}\ and\ \bibinfo
  {author} {\bibfnamefont{S.-C.}\ \bibnamefont{Zhang}},\ }%
  \bibfield{journal}{%
  \bibinfo {journal} {Phys. Status Solidi Rapid Res. Lett.}\ }%
  \textbf{\bibinfo {volume} {7}},\ \bibinfo {pages} {72} (\bibinfo {year}
  {2013})%
  \bibAnnoteFile{NoStop}{zhang2013TI}%
\bibitem{haldane1988model}%
  \BibitemOpen
  \bibfield{author}{%
  \bibinfo {author} {\bibfnamefont{F.~D.~M.}\ \bibnamefont{Haldane}},\ }%
  \bibfield{journal}{%
  \bibinfo {journal} {Phys. Rev. Lett.}\ }%
  \textbf{\bibinfo {volume} {61}},\ \bibinfo {pages} {2015} (\bibinfo {year}
  {1988})%
  \bibAnnoteFile{NoStop}{haldane1988model}%
\bibitem{yu2010quantized}%
  \BibitemOpen
  \bibfield{author}{%
  \bibinfo {author} {\bibfnamefont{R.}~\bibnamefont{Yu}}, \bibinfo {author}
  {\bibfnamefont{W.}~\bibnamefont{Zhang}}, \bibinfo {author}
  {\bibfnamefont{H.-J.}\ \bibnamefont{Zhang}}, \bibinfo {author}
  {\bibfnamefont{S.-C.}\ \bibnamefont{Zhang}}, \bibinfo {author}
  {\bibfnamefont{X.}~\bibnamefont{Dai}},\ and\ \bibinfo {author}
  {\bibfnamefont{Z.}~\bibnamefont{Fang}},\ }%
  \bibfield{journal}{%
  \bibinfo {journal} {Science}\ }%
  \textbf{\bibinfo {volume} {329}},\ \bibinfo {pages} {61} (\bibinfo {year}
  {2010})%
  \bibAnnoteFile{NoStop}{yu2010quantized}%
\bibitem{chang2013experimental}%
  \BibitemOpen
  \bibfield{author}{%
  \bibinfo {author} {\bibfnamefont{C.-Z.}\ \bibnamefont{Chang}}, \bibinfo
  {author} {\bibfnamefont{J.}~\bibnamefont{Zhang}}, \bibinfo {author}
  {\bibfnamefont{X.}~\bibnamefont{Feng}}, \bibinfo {author}
  {\bibfnamefont{J.}~\bibnamefont{Shen}}, \bibinfo {author}
  {\bibfnamefont{Z.}~\bibnamefont{Zhang}}, \bibinfo {author}
  {\bibfnamefont{M.}~\bibnamefont{Guo}}, \bibinfo {author}
  {\bibfnamefont{K.}~\bibnamefont{Li}}, \bibinfo {author}
  {\bibfnamefont{Y.}~\bibnamefont{Ou}}, \bibinfo {author}
  {\bibfnamefont{P.}~\bibnamefont{Wei}}, \bibinfo {author}
  {\bibfnamefont{L.-L.}\ \bibnamefont{Wang}}, \emph{et~al.},\ }%
  \bibfield{journal}{%
  \bibinfo {journal} {Science}\ }%
  \textbf{\bibinfo {volume} {340}},\ \bibinfo {pages} {167} (\bibinfo {year}
  {2013})%
  \bibAnnoteFile{NoStop}{chang2013experimental}%
\bibitem{zhang2013topology}%
  \BibitemOpen
  \bibfield{author}{%
  \bibinfo {author} {\bibfnamefont{J.}~\bibnamefont{Zhang}}, \bibinfo {author}
  {\bibfnamefont{C.-Z.}\ \bibnamefont{Chang}}, \bibinfo {author}
  {\bibfnamefont{P.}~\bibnamefont{Tang}}, \bibinfo {author}
  {\bibfnamefont{Z.}~\bibnamefont{Zhang}}, \bibinfo {author}
  {\bibfnamefont{X.}~\bibnamefont{Feng}}, \bibinfo {author}
  {\bibfnamefont{K.}~\bibnamefont{Li}}, \bibinfo {author}
  {\bibfnamefont{L.-l.}\ \bibnamefont{Wang}}, \bibinfo {author}
  {\bibfnamefont{X.}~\bibnamefont{Chen}}, \bibinfo {author}
  {\bibfnamefont{C.}~\bibnamefont{Liu}}, \bibinfo {author}
  {\bibfnamefont{W.}~\bibnamefont{Duan}}, \emph{et~al.},\ }%
  \bibfield{journal}{%
  \bibinfo {journal} {Science}\ }%
  \textbf{\bibinfo {volume} {339}},\ \bibinfo {pages} {1582} (\bibinfo {year}
  {2013})%
  \bibAnnoteFile{NoStop}{zhang2013topology}%
\end{thebibliography}
%

\end{document}